\documentclass{aa}
\usepackage{epsf}
\begin{document}
\title{Radio observations of new galactic bulge planetary nebulae}
\subtitle{}
\author{G.C. Van de Steene\inst{1,3} 
\and G.H. Jacoby\inst{2,4} \thanks{Based on data acquired at the Australia Telescope Compact Array.
The Australia Telescope is funded by the Commonwealth of Australia for
operation as a National Facility managed by CSIRO. }}

\institute{ Royal Observatory of Belgium, Ringlaan 3, 1180 Brussels, Belgium
\and 
National Optical Astronomical Observatories, P.O. Box 26732, Tucson, AZ 85726, U.S.A.
\and
Research School for Astronomy and Astrophysics, Private Bag P.O., Weston Creek,
ACT 2611, Australia
\and WIYN Observatory, P.O. Box 26732, Tucson, AZ 85726, U.S.A.
}

\offprints{G.C. Van de Steene\inst{1},\email{gsteene@oma.be}}

\date{received / accepted}

\abstract{
We observed 64 newly identified galactic bulge planetary nebulae in
the radio continuum at 3 and 6~cm with the Australia Telescope Compact
Array.  We present their radio images, positions, flux densities, and
angular sizes.  The survey appears to have detected a larger ratio of
more extended planetary nebulae with low surface brightness than in
previous surveys.  We calculated their distances according to Van de
Steene \& Zijlstra (\cite{VdSteene95}). We find that most of the new sample 
is located on the near side around the galactic center and 
closer in than the previously known bulge PNe. Based on H$\alpha$ images and
spectroscopic data, we calculated the total H$\alpha$ flux. We compare
this flux value with the radio flux density and derive the extinction.
We confirm that the distribution of the extinction values around the
galactic center rises toward the center, as expected.
\keywords{planetary nebulae: general, galactic bulge, radio}
}
 
\titlerunning{Radio observations of new GB PNe}
\authorrunning{Van de Steene \&  Jacoby}

\maketitle

\section{Introduction}

Planetary Nebulae (PNe) are bright emission line objects, observable
throughout the Galaxy.  They are excellent probes of abundance
gradients, the chemical enrichment history of the interstellar medium, 
the effects of metallicity on stellar evolution, and kinematics.

Most small PNe ($\sim$90$\%$) within 10\degr\ of the galactic center
are physically close to it (Pottasch \& Acker 1989).  Since they can
be assumed to be at the same known distance of $\sim$7.8 kpc, their
distance-dependent parameters, such as luminosity and size, can be
determined. These parameters are needed to define the underlying
population.  The chemical composition and the central star parameters
are computed via self-consistent photo-ionization modelling of the
nebula.  Because luminosity correlates with the central star mass,
which correlates with the progenitor mass, which, in turn, correlates
with stellar age, the relationship between age and composition can be
deduced (Dopita et al. \cite{Dopita97}, Walsh et
al. \cite{Walsh00}). The chemical enrichment history of the Bulge
could be tracked using PNe.

We surveyed a 4~x~4~degree field centered on the galactic center in
[S\,{\sc iii}]$\lambda$9532 and a continuum band at KPNO with the
60-cm Schmidt telescope and a 2048~x~2048 pixels thick STIS CCD in
July 1994 and June 1995.  The field of view was
65\arcmin\,x\,65\arcmin\ and the pixel size 2\arcsec.  This survey has
uncovered 95 new PN candidates in addition to the 34 previously known
in this region (Acker et al. 1992, Kohoutek 1994).  45 PNe were
confirmed via optical spectroscopy with the 1.52-m ESO telescope and
the Boller \& Chivens spectrograph, while 19 fainter ones were
confirmed at the CTIO 4-m with the RC spectrograph (Van de Steene \&
Jacoby 2001, in preparation).

Accurate radio flux densities and angular diameters are crucial to obtain a
good photo-ionization model of the PNe (van Hoof \& Van de Steene
\cite{vHoof99}).  The very high extinction causes the H$\beta$ line to be faint
or even undetected in the optical spectra. Hence the radio flux density is
needed to determine the extinction and the total ionizing flux.

In this article we present the radio continuum observations of 64 PNe
confirmed spectroscopically with the ESO 1.52-m and CTIO 4-m telescopes.
We describe the observations in Sect. \ref{observations} and the
data reduction in Sect. \ref{reduction}.  The results are presented
in Sect. \ref{results}. The improved method for determining the
distances, based on a relationship between radii and
radio surface brightness (Van de Steene \& Zijlstra \cite{VdSteene95}) is used in
Sect. \ref{distances} to determine the distances of the PNe 
and discuss their distribution in the galactic bulge.
In Sect. \ref{extinction} we determine the extinction values
of these new bulge PNe.

\section{Observations}
\label{observations}

Radio continuum observations for 64 new PNe which had been confirmed
spectroscopically at the ESO 1.52-m and CTIO 4-m telescopes were
obtained with the Australia Telescope Compact Array. 
The array was in configuration $\#$ 6A.
The shortest baseline was 337 m and the longest 5939 m.  The bandwidth
was 128 MHz
centered at 4800~MHz and 8640~MHz, corresponding to 6 and 3~cm
respectively.  The synthesized beam for the 6 km array at 6 cm is
$\sim$ 2\arcsec\ and $\sim$1\arcsec\ at 3 cm.  According to the ACA
manual, the largest well imaged structure in a 60~min observation is
30\arcsec\ at 6 cm and 15\arcsec\ at 3 cm. Consequently we expected
all our sources to be well imaged at 6 cm and virtually all at 3 cm.
Forty PNe were
observed at 3 and 6~cm simultaneously for 12~h on each of 12, 13, 16 \& 17
February 1997 and 36 PNe were observed during 10~h on each of 24, 25, and 26
May 1999. 
In 1997 each subsample contained 10 PNe and each PN was observed for 
5~min every hour.  In 1999 each subsample contained 6 PNe and each PN
was observed for 8~min every hour.  The total quality integration time
per PN was at least 50~min in 1997 and 80~min in 1999.
To avoid artifacts at the center of the field we offsetted the
positions 30\arcsec\ in declination.  At the beginning and the end of
each 12~h shift the primary flux density calibrator 1934-638 was
observed for 5 to 10~min.  The phase calibrator VLA~1730-130 and 
VLA~1748-253 were observed about every 20~min for 2~min.

\section{Data Reduction}
\label{reduction}

The data were reduced using the package {\tt Miriad} following
standard reduction steps as described in the reference guide by Bob
Sault and Neil Killeen (Sault et al. 1995) . The data were loaded, bad
points flagged, and then split into single source files.  
The bandpass function and the antenna gains were determined as
a function of time to calibrate the data.  
From the visibilities images were made using the multi-frequency
synthesis technique and robust weighting with a robustness parameter
of 0.5, which gives nearly the same sensitivity as natural weighting
but with a significantly better beam. First a low resolution image of
the primary beam was made to identify confusing sources.  Next a high
resolution image was made including all sources. These maps were {\tt
CLEAN}ed.
Because we had offsetted the PNe
from the field center, we applied the primary beam correction which
amounted to about 1\% at 6 cm and 2\% at 3 cm.  
Self-calibration was not applied because the
flux values of the PNe are too low. Contour plots of the detected PNe are
presented in the Appendix.

\section{Analysis and Results}
\label{results}

\subsection{Positions}

In Table \ref{tabpos} we list the position of the peak flux density per beam 
of the PN at 6~cm
together with the optical positions as determined from the H$\alpha$
or [S III]$\lambda$9532 images (Jacoby \& Van de Steene, 2001, in preparation).
The peak of the radio emission is adopted as the PN position.
If the PN is extended in the radio, the radio position may be off
center.  The radio position will be better determined for higher peak
flux values and smaller PNe.  It is also for this reason that we chose to
use the value at 6~cm and not at 3~cm, where the resolution is twice
as high and the signal to noise per beam lower for extended sources.

We note that the radio positions have a tendency to be offset
towards the west of the optical position. In declination there is no
clear tendency noticeable in the offsets.

The optical and radio positions agree very well.  PNe for which the
radio position differs more than 2 \arcsec\ in RA or DEC from the
optical position are extended and the peak  in the radio is
usually not centered.

Of the 64 PNe observed,  7 were not detected: 
 JaSt~7, 21, 45, 80, 88, 92, and 96.  Most likely 
they are very extended and have a surface
brightness that is too low to be detected in the radio.
They are also very faint in the H$\alpha$ images and
their H$\alpha$ flux values are very uncertain (Jacoby \& Van de Steene
2001, in preparation). All but JaSt 96 were also faint and extended
in the [S III]$\lambda$9532 images.

\begin{table*}
\caption[]{Optical and radio positions of the  new galactic bulge PNe. 
The radio position is the position of the peak flux density per beam of the PN at 6~cm. 
Objects for which run=1 were observed
in 1997, objects for which run=2 were observed in 1999 }
\begin{tabular}{lllllllll}
\hline
JaSt & \multicolumn{2}{l}{Optical (2000.0)} & \multicolumn{2}{l}{Radio (2000.0)} & $\Delta$RA & $\Delta$DEC  & run & comment \\
     & RA(h m  s) & DEC (\degr\ \arcmin\ \arcsec)  &  RA(h  m  s) & DEC (\degr\ \arcmin\ \arcsec) & \arcsec\ & \arcsec\ &  & \\
\hline
1  & 17 34 43.64 &  -29 47 05.03 &  17 34 43.60 & -29 47 06.03 & -0.60  &  1.00    & 2 &   \\  
 2 & 17 35 00.96 &  -29 22 15.72 &  17 35 00.96 & -29 22 15.85 &   0.00 &  0.13    & 1 &  \\  
 3 & 17 35 22.90 &  -29 22 17.58 &  17 35 22.79 & -29 22 17.08 & -1.65 & -0.50   & 1 &  \\ 
 4 & 17 35 37.47 &  -29 13 17.67 &  17 35 37.39 & -29 13 17.81 & -1.20 & 0.14     & 1 & \\ 
 5 & 17 35 52.51 &  -28 58 27.95 &  17 35 52.44 & -28 58 27.45 & -1.05 &  -0.50  & 1 & \\ 
7  & 17 38 26.69 &  -28 47 06.48 &   &  &   &     & 2 &  not detected \\  
 8 & 17 38 27.74 &  -28 52 01.31 &  17 38 27.58 & -28 52 01.36 &  -2.40 & 0.05  & 2 &  \\ 
 9 & 17 38 45.64 &  -29 08 59.27 &  17 38 45.54 & -29 08 55.65 &  -1.50  &  -3.62 & 2 &  \\  
11 & 17 39 00.55 &  -30 11 35.23 & 17 39 00.48 & -30 11 32.23 &  -1.05 &-3.00  & 2 &  \\  
16 & 17 39 22.70 &  -29 41 46.08 &  17 39 22.64 & -29 41 45.35 &  -0.90 &-0.70  & 1 &  \\  
17 & 17 39 31.32 &  -27 27 46.78 &  17 39 31.21 & -27 27 47.28 &  -1.65 & -0.50  & 1 & \\  
19 & 17 39 39.38 &  -27 47 22.58 &  17 39 39.31 & -27 47 22.21 &  -1.05 & -0.37  & 1 &  \\  
21 & 17 39 52.92 &  -27 44 20.54 &                      &                      &           &             & 2 & not detected \\  
23 & 17 40 23.17 &  -27 49 12.04 &  17 40 23.08 & -27 49 12.29 &  -1.35 & 0.25  & 1 & \\  
24 & 17 40 28.23 &  -30 13 51.30 &  17 40 28.19 & -30 13 51.00 &  -0.60 & -0.30 &  1 &  \\ 
26 & 17 40 33.52 &  -29 46 14.98 &  17 40 33.25 & -29 46 12.48 &  -4.05 & -2.50 &  2 &   \\  
27 & 17 40 42.34 &  -28 12 31.90 &  17 40  42.16 & -28 12 30.81 & -2.7 & -1.1 & 1 & \\
31 & 17 41 27.93 &  -28 52 51.61 &  17 41 27.89 & -28 52 50.61 & -0.60 & -1.00 & 1 & \\   
34 & 17 41 54.80 &  -27 03 20.33 &  17 41 54.69 & -27 03 18.33 & -1.65 & -2.00 & 2 & \\ 
36 & 17 42 25.20 &  -27 55 36.36 &  17 42 25.13 & -27 55 36.36 & -1.05 & 0.00 & 1 & \\  
37 & 17 42 28.60 &  -30 09 34.93 &  17 42 28.52 & -30 09 32.93 & -2.00 & -2.00 & 1 & \\  
38 & 17 42 32.41 &  -27 33 15.18 &  17 42 32.29 & -27 33 16.26 & -1.80 &1.08 & 1 & \\  
41 & 17 42 49.96 &  -27 21 19.68 &  17 42 49.85 & -27 21 19.31 & -1.65 & -0.37 & 1 &  \\  
42 & 17 43 17.06 &  -26 44 17.67 &  17 43 16.98 & -26 44 18.25 & -1.20 & 0.58 & 1 & \\  
44 & 17 43 23.48 &  -27 34 06.03 &  17 43 23.52 & -27 34 06.55 &  0.60 & 0.52 & 1 & \\  
45 & 17 43 23.44 &  -27 11 16.91 &                     &             &          &         & 2 &  not detected \\  
46 & 17 43 30.43 &  -26 47 32.33 &  17 43 30.36 & -26 47 31.83 & -1.05 & -0.50 & 1 & \\  
49 & 17 44 04.34 &  -28 15 57.86 &  17 44 04.23 & -28 15 56.99 & -1.65 & -0.87 & 1  & \\  
52 & 17 44 37.30 &  -26 47 25.23 &  17 44 37.23 & -26 47 25.23 & -1.05 & 0.00 & 1 & \\ 
54 & 17 45 11.06 &  -27 32 36.80 &  17 45 11.02 & -27 32 36.80 & -0.60 & 0.00 & 2 & \\  
55 & 17 45 37.36 &  -27 01 18.44 &  17 45 37.15 & -27 01 18.16 & -3.15 & -0.28 & 1 &  \\  
56 & 17 45 47.05 &  -27 30 42.03 &  17 45 46.96 & -27 30 42.01 & -1.35 &-0.02 & 2 & \\  
58 & 17 46 52.20 &  -30 37 42.83 &  17 46 52.32 & -30 37 42.33 & 1.80  & -0.50 & 2 &  \\  
60 & 17 47 53.91 &  -29 36 49.67 &  17 47 53.91 & -29 36 49.67 & 0.00 & 0.00 &  2 & \\ 
63 & 17 48 46.27 &  -27 25 37.22 &  17 48 46.27 & -27 25 36.72 & 0.00 & -0.50 & 1  & \\   
64 & 17 48 56.04 &  -31 06 41.95 &  17 48 56.04 & -31 06 42.45 & 0.00 & 0.50 & 1  & \\  
65 & 17 49 20.02 &  -30 36 05.57  & 17 49 20.02 &  -30 36 05.08 & 0.00 & 0.23 &  1   & \\  
66 & 17 49 22.15 &  -29 59 27.02  & 17 49 22.11 & -29 59 27.02 &  -0.60 & 0.00  & 2 & \\ 
67 & 17 49 28.10 &  -29 20 47.56  & 17 49 28.02 & -29 20 47.56 & -1.20 & 0.00   &  2 & \\ 
68 & 17 49 50.87 &  -30 03 10.47  & 17 49 50.83 & -30 03 10.97 & -0.60 & 0.50   & 2 &  \\  
69 & 17 50 10.04 &  -29 19 05.14  &  17 50 10.11: & -29 19 08.82: & 1.05 & 3.68 & 1 &  \\  
70 & 17 50 21.07 &  -28 39 02.46 &  17 50 21.03 & -28 39 01.46 &  -0.60 & 1.0 & 2 & \\ 
71 & 17 50 23.32 &  -28 33 10.95  &  17 50 23.21 & -28 33 10.45 & -1.65 & -0.50 & 1 & \\ 
73 & 17 50 47.74 &  -29 53 16.01  &  17 50 47.82 &  -29 53 14.01  &  1.20 & -1.99 & 2 & \\ 
74 & 17 50 46.85 & -28 44 35.42  &  17 50 46.77 & -28 44 34.92 & -1.20 & -0.50 & 1 & \\ 
75 & 17 50 48.08 &  -29 24 43.60 &  17 50 48.00 & -29 24 43.18 & -1.20 & -0.42 & 1 & \\ 
76 & 17 50 56.47 &  -28 31 24.63 & 17 50 56.33 &  -28 31 24.13 & -2.10 &-0.50 & 2 & \\  
77 & 17 51 11.65 & -28 56 27.20  &  17 51 11.58 & -28 56 27.20 & -1.20 & 0.00 & 1 & \\ 
78 & 17 51 24.68 &  -28 35 40.34 &  17 51 24.79 & -28 35 39.44 & 1.65 & -0.90 & 1 & \\ 
79 & 17 51 53.63 & -29 30 53.41  &  17 51 53.55 & -29 30 53.41 & -1.20 & 0.00 & 1 & \\ 
80 & 17 51 55.54  &  -27 48 02.46 &                    &  &  &  & 1 & not detected \\ 
81 & 17 52 04.35 & -27 36 39.28  &  17 52 04.28 & -27 36 38.28 & -1.05 & 1.00 & 1 & \\ 
83 & 17 52 45.17 &  -29 51 05.21 &  17 52 45.17 & -29 51 03.30 & 0.00 & -1.91 & 1 & \\ 
85 & 17 52 49.05 &  -29 41 54.92 & 17 52 48.97  &  -29 41 55.92 & -1.20 & 1.0 & 2 & \\ 
\hline
\end{tabular}
\label{tabpos}
\end{table*}
\addtocounter{table}{-1}
\begin{table*}
\caption[]{Table \ref{tabpos} continued  }
\begin{tabular}{lllllllll}
\hline
JaSt & \multicolumn{2}{l}{Optical (2000.0)} & \multicolumn{2}{l}{Radio (2000.0)} & $\Delta$RA & $\Delta$DEC  & run & comment \\
     & RA(h m  s) & DEC (\degr\ \arcmin\ \arcsec)  &  RA(h  m  s) & DEC (\degr\ \arcmin\ \arcsec) & \arcsec\ & \arcsec\ &  & \\
\hline
86 & 17 52 52.20 &  -29 30 00.07 &  17 52 52.16 & -29 30 01.08 & -0.60 & 1.08 & 1 &  \\ 
88  & 17 53 00.89  & -29 05 44.08  &  & & &  & 2 & not detected   \\  
89  & 17 53 06.67 & -28 18 07.91  &  17 53 06.79 & -28 18 10.47 & 1.80 & 2.56 & 1&    \\ 
90 &  17 53 17.77  & -28 04 33.20 & 17 53 17.73 &  -28 04 32.67 &  -0.60 & -0.53 & 2 &  \\ 
92 & 17 53 19.81 &  -28 27 14.67 &    & & &  & 2 & not detected  \\ 
93 & 17 53 24.14 &  -29 49 48.45 &  17 53 24.05: & -29 49 51.15: & -1.35 & 2.70 & 1 &  \\ 
95 & 17 53 35.38 & -28 28 51.02  &  17 53 35.51 & -28 28 51.96 & 1.95 & 0.94 & 1 & \\  
96 & 17 53 57.16 &  -29 20 14.97 &  & & &  & 2 &  not detected \\  
97 & 17 54 13.36 &  -28 05 16.82 &  17 54 13.28 & -28 05 16.82 &  -1.20 & 0.00 & 1 & \\ 
98 & 17 55 46.39 & -27 53 38.91  &  17 55 46.28 & -27 53 38.41 & -1.65 & -0.50 & 1 &  \\ 
\hline
\end{tabular}
\end{table*}



\small
\begin{table*}
\caption[]
{Radio flux densities at 6 and 3 cm, and angular sizes are presented.  
The name of the PN is listed in the first
column. The second to sixth columns give the peak flux density per beam, 
the total radio flux density, the error estimate, and the rms in the map respectively.
Columns six to nine list the same quantities at 3~cm.  The tenth
column gives the deconvolved FWHM in arcsec ($\Delta\delta$ x
$\Delta\alpha$). The eleventh column gives the FWHM of the beam major
and minor axis in arcsec ($\Delta\delta$ x $\Delta\alpha$).  The
thirteenth column gives the equivalent FWHM diameter. The fourteenth
column indicates whether the diameter FWHM is the deconvolved FWHM
determined after fitting a Gaussian model (G) or after measuring the
size from the contour plot (C).  All diameters were determined at 6~cm
unless otherwise mentioned. The last column gives an estimate of the
full equivalent diameter as explained in the text.  PS stands for
Point Source.  Values which are very uncertain are marked with a
colon.}
\begin{tabular}{lllllllllllllll}
\hline
Name & 6cm    &      &     &     & 3cm     &      &     &    & size &  beam & FWHM & method  & $\theta$ \\ 
 JaSt  & Peak & Flux & error & noise &  Peak & Flux & error & noise & $\Delta\delta$ x $\Delta\alpha$ & maj x min  &  diam &  & \\ 
     & mJy/ & mJy & mJy & mJy/ & mJy/ & mJy & mJy & mJy/ & \arcsec\ x \arcsec\ & \arcsec\ x \arcsec\ & \arcsec\ & & \\ 
     & beam &   &    &  & beam & beam & & & beam & & & & & \\
\hline
1 &  0.85 & 1.6 &  0.2 &  0.10 & 0.47 & 1.4 & 0.6 & 0.12 &  4.5 x  2.6  &  4.3  x 1.9 & 3.4 & G & 5.5   \\ 
2  & 1.8  & 4.3  & 0.5 & 0.22  & 0.82  & 4.2 &  1.4 & 0.12 & 5.0 x 4.0 & 4.7 x 1.8 & 4.5  & C  & 6.3 \\ 
3  & 4.0  & 12.4 & 1.2 & 0.17  & 1.6   & 10.0 & 1.0 & 0.13 & 5.8 x 4.4  & 4.8 x 2.0 & 5.1 &  C   &  9.2 \\ 
4  & 1.2  & 4.1  &  0.4 & 0.20  & 0.65  & 3.5 & 1.2  & 0.15 & 7.7 x 7.1  & 4.7 x 1.8 & 7.4 & C  & 11.2  \\ 
5  & 4.0  & 11.3 & 1.1 & 0.15  & 1.8   & 10.5 & 1.0 & 0.15 & 6.2 x 5.1  & 4.9 x 2.0 & 5.6 & C  & 8.4  \\ 
8  &1.2 &    3.8 &  0.4 & 0.08 & 0.5 &  2.6 &  0.6 & 0.04 & 3.6 x 5.8 & 4.4 x 1.9 & 4.6 & C  & 7.2 \\ 
9  & 0.96 &  4.0 &   0.4 & 0.08 & 0.38 &  2.5 & 0.8 &  0.08 &  5.6 x 6.2 &  4.7 x 1.8 & 5.9 & C & 8.1 \\ 
11 & 0.76  & 3.3 &  0.3 & 0.07 & 0.41 &  $>$2.0 &  & 0.12 & 6.9 x 5.9 & 4.6 x 1.8 & 6.4 & C &  10.9  \\  
16 & 13.6 & 27.0 & 2.8 & 0.27  & 7.2   & 24.7 & 2.5 & 0.30 & 3.4 x 2.4  & 4.8 x 2.0 & 2.9 & G  & 4.8 \\ 
17 & 2.5  & 9.4 & 1.0 & 0.15  & 1.2   & 8.2  & 0.9 & 0.13 &  5.4 x 5.3  & 5.7 x 1.8 & 5.3 & C   & 7.9 \\ 
19 & 2.1  & 6.8  & 0.8 & 0.29  & 1.0   & 5.5  & 0.6 & 0.14 & 3.6 x 3.7  & 5.3 x 1.6 &  3.6  & C & 6.8 \\ 
23 & 3.3  & 3.7  & 0.4 & 0.25  & 2.6   & 3.9  & 0.4 & 0.21 & PS &  3.1 x 1.1 & $<$1.8 & G$_{3cm}$ &  \\ 
24 & 6.3  & 16.0 & 1.7 & 0.27  & 2.7   & 14.1 & 1.4 & 0.23 & 4.8 x 2.7  & 4.9 x 1.7  & 3.6 & G & 5.8 \\ 
26 &  2.4 &  14.1 &   1.4 & 0.09 & 1.0 &  9.0 &  0.9 & 0.08 & 9.4 x 9.3  & 5.1 x  1.7 & 8.5 & C & 11.1\\ 
27 & 0.8 & 2.5 & 0.3 & 0.14 & 0.5 &    &       & 0.15 & 9.9 x 5.3 & 5.3 x 1.7 & 8.2 & C & 8.7 \\
31 & 3.3  & 11.5 & 1.2 & 0.22  & 2.4   & $>$3.0  &  & 0.61 & 5.0 x 5.1 &  5.1 x 1.7 & 5.0 & C & 9.0 \\ 
34 &  0.58 &  1.7 & 0.2 & 0.07 & 0.34 & $>$1.5 &    &  0.09 &  3.2 x 6.0  & 5.1 x 1.7 & 4.4  & C & 7.1 \\ 
36 & 18.6 & 31.1 & 3.1 & 0.33  & 11.7  & 31.1 & 3.1 & 0.35 &  2.6 x 1.9  &  3.1 x 1.1 & 2.2 & G$_{3cm}$ &  3.6  \\ 
37 & 5.2  & 16.0 & 1.7 & 0.70  & 2.5   &  $>$5.5  &  & 0.35 & 5.7 x 3.8  &  4.6 x 1.5 & 4.6 & C & 7.4  \\ 
38 & 1.2  & 3.0  & 0.6 & 0.28  & 0.9   & 2.0 & 0.7& 0.15 & 6.8 x 2.2 &  5.3 x 1.7 & 3.9 & C & 7.0  \\ 
41 & 8.0  & 16.7 & 1.7 & 0.20  & 4.6   & 17.6 & 1.8 & 0.18 & 3.4 x 3.2  &  5.6 x 1.8 & 3.3 & C & 6.3 \\ 
42 & 5.6  & 12.9 & 1.3 & 0.21  & 3.3   & 15.5 & 1.6 & 0.26 & 4.0 x 2.5   &  5.5 x 1.6 & 3.1 & G & 5.0 \\ 
44 & 1.3  & 3.8 & 1.0 & 0.27  & 0.9   & $>$1.8  &  & 0.27 & 12.2: x 6.0: &  5.4 x 1.6 & 8.6: & C  & 8.6:  \\ 
46 & 14.6 & 20.8 & 2.1 & 0.40  & 12.3  & 20.5 & 2.1 & 0.30 &  2.0 x 1.1   &  3.3 x 1.1 & 1.5  &  G    &  2.5 \\ 
49 & 3.1  & 8.2  & 0.9 & 0.27  & 2.1   & 10.8 & 1.1 & 0.21 & 4.6 x 3.6  &  5.2 x 1.6 & 4.0  &  C    &  6.6  \\ 
52 & 20.8 & 24.0 & 2.4 & 0.34 & 16.9  & 22.8 & 2.3 & 0.40 & 2.5 x 0.4   &  3.3 x 1.1 & 1.0  &  G$_{3cm}$ &  1.7 \\ 
54 & 21.9  &  82.3 & 8.2 & 0.08 & 7.5 &   75.2 &   7.5 &  0.09 &  4.7 x 4.3 &  5.1 x 1.8 & 4.5 & G & 7.2 \\    
55 & 2.0  & 9.2 & 0.9 & 0.19  & 0.9   & 10.0  & 2.0 & 0.17 & 5.5 x 6.4   & 5.4 x 1.6  & 6.0 &  C   &  10.7 \\ 
56 & 5.8  & 14.1 & 1.4  & 0.09  & 2.4 &12.2 &  1.2 & 0.07 & 3.8 x  2.7 &  2.8 x  .95 & 3.2 & G$_{3cm}$  & 5.0  \\ 
58 & 3.5 & 29.5 &  2.9 & 0.06 & 1.3 & 21.3 &  2.1  & 0.07 & 8.9 x 8.0  & 4.6 x 1.7  & 8.4 & C & 15.5 \\  
60 & 6.1 & 6.5 & 0.7 & 0.06 & 7.0 &  7.9 &   0.8 &  0.07 &  .85  x   .28 &  2.6 x 1.0 & 0.5 & G$_{3cm}$ & 0.9 \\ 
63 & 1.1  & 3.9  & 0.4 & 0.11  & 0.7   & 3.6 & 1.2  & 0.13 & 6.1 x 6.0  & 5.5 x 1.8  & 6.0  &  C   & 8.4 \\ 
64 & 23.8 & 28.3 & 2.9 & 0.42  & 15.8  & 25.6 & 2.6 & 0.58  & 2.0 x 0.7   & 5.0 x 1.9  & 1.4  &  G    & 2.4     \\ 
65 & 3.3 & 3.4 & 0.4 & 0.23 & 6.3 & 6.6 & 0.7 & 0.30 & PS & 2.7 x 1.0 & $<$ 1.6 & G$_{3cm}$ & \\  
66 & 45.9 &  65.3 & 6.5 &  0.15 & 26.0 & 60.9 & 6.1 &  0.10 & 2.0 x 1.5 &  2.6 x 1.00 & 1.7 & G$_{3cm}$ & 2.8  \\ 
67 & 24.6 &   27.8 &  2.8 &  0.08 & 17.7 &  26.0 &   2.6  & 0.10 & 1.3 x .9 & 2.5 x 1.0 &  1.1 & G $_{3cm}$ & 1.8 \\
68 & 5.4 & 5.9 & 0.6 & 0.06 & 7.7 & 8.8 & 0.9 &  0.10 & .8 x .3 & 2.5 x 1.0 & 1.5 &  G$_{3cm}$  & 0.9 \\ 
69 & 0.9  & $>$2.0  &  & 0.26  & 0.5  &  $>$2.0  &  & 0.15 &  &    4.9 x 1.7  &  &   &   \\ 
70 & 4.0 &  13.0 &  1.3 & 0.11 & 1.9 & 11.5 & 1.2 & 0.06 & 4.3 x 5.3 &  4.6  x  1.7 & 4.8 & C &  8.5 \\ 
71 & 20.2 & 34.0 & 3.5 & 0.50  & 10.0  & 36.0 & 3.6 & 0.55 & 4.6 x 1.6  &  5.4 x 1.9  &  2.7 & G &  4.5  \\ 
73 &11.7 & 13.7 & 1.4 & 0.09 & 9.4 & 13.0 & 1.3 &  0.11 & 1.3 x .7  & 2.6 x 1.0 & 0.95 & G$_{3cm}$ & 1.6 \\  
74 & 12.7 & 22.7 & 2.3 & 0.28  & 6.1   & 22.2 & 2.2 & 0.25 &  4.5 x 2.2  &  3.3 x 1.0  & 3.1 & G$_{3cm}$ &  4.9  \\ 
75 & 16.8 & 22.5 & 2.3 & 0.53  & 11.0  & 19.0 & 2.0 & 0.55 & 3.0 x 0.9  &  5.0 x 1.6  & 1.6  & G & 2.7 \\ 
76 & 5.4 &  7.4 &  0.7 & 0.07 & 2.5 & 6.7 & 0.7 & 0.08 & 3.9 x 1.8  & 2.4 x 1.0 & 2.6 & G$_{3cm}$ & 4.7 \\ 
77 & 39.4 & 49.0 & 5.0 & 0.85  & 34.0  & 51.2 & 5.1 & 0.98 & PS &  3.1 x 1.1 & $<$1.5  & G$_{3cm}$  &  \\ 
78 & 1.5  & 12.3 & 1.2 & 0.18  & 0.9   & 9.0  & 2.3 & 0.15 & 12.0 x 9.0   &  5.1 x 1.6 & 10.4 & C & 14.6 \\ 
79 & 4.1  & 4.1  & 0.4 & 0.14  & 6.7   & 7.0  & 0.7 & 0.23 &  PS & 2.7 x 1.2  & $<$1.5  & G$_{3cm}$ &  \\ 
\hline
\end{tabular}
\label{tabflux}
\end{table*}
\addtocounter{table}{-1}
\begin{table*}
\caption[]
{ Table \ref{tabflux} continued}
\begin{tabular}{lllllllllllllll}
\hline
Name & 6cm    &      &     &     & 3cm     &      &     &    & size &  beam & FWHM & method  & $\theta$ \\ 
 JaSt  & Peak & Flux & error & noise &  Peak & Flux & error & noise & $\Delta\delta$ x $\Delta\alpha$ & maj x min  &  diam &  & \\ 
     & mJy/ & mJy & mJy & mJy/ & mJy/ & mJy & mJy & mJy/ & \arcsec\ x \arcsec\ & \arcsec\ x \arcsec\ & \arcsec\ & & \\ 
     & beam &   &    &  & beam & beam & & & beam & & & & & \\
\hline
81 & 17.8 & 20.8 & 2.0 & 0.25   & 14.1  & 20.8 & 2.1 & 0.29 & 1.8 x 0.7    & 5.1 x 2.0 & 1.3  &  G  & 2.2 \\ 
83 & 1.7  & 5.0:  & 2.5 & 0.40  & 0.6   & $>$0.6  &  & 0.16 & 5.3: x 5.7:   &  4.5 x 1.8 & $>$5.5: &  C & $>$6.3 \\ 
85 & 0.6 &  4.4 & 1.2 & 0.09 &  &  &  & & 13.1: x 13.1: & 4.2 x 1.9 & 13.1: & C & 17.5: \\ 
86 & 1.7  & 8.5  & 0.9 & 0.17  & 0.9   & $>$2.5 &  & 0.17 &  7.5 x 7.7  &  4.7 x 1.8 & 7.6 & C  & 11.4 \\ 
89 & 1.8  & 8.5  & 0.9 & 0.27   & 0.8   & $>$2.0  &  & 0.19 & 9.5 x 5.5    &  4.8 x 1.7 & 7.2 & C & 8.6  \\ 
90 & 0.51& 1.0 & 0.2 & 0.07 &  & & & 0.13 & 4.5  x  2.5 &  4.4  x  1.9 & 3.4 & G & 5.5 \\ 
93 & 0.8  & 6.3  & 0.9 & 0.15   & 0.6   & $>$2.1  &  & 0.13 & 20.2 x 8.0   &  4.8 x 2.0 & 12.7: & C & 17.8: \\ 
95 & 1.1  & 5.5  & 1.0 & 0.25   & 0.7   & 4.5:  & 1.5 & 0.18 & 14.6  x 7.5   &   4.8 x 1.8  &  10.5 & C & 12.6 \\ 
97 & 2.7  & 11.5 & 1.2 & 0.16   & 1.2   & 11.8 & 1.2 & 0.18 & 8.0 x 5.9  &  5.2 x 1.9 & 6.8 & C & 11.4 \\ 
98 & 21.0 & 21.8 & 2.2 & 0.38   & 20.3  & 24.0 & 2.4 & 0.46 & PS &  2.9 x 1.1 & $<$1.5 & G$_{3cm}$ &  \\ 
\hline
\end{tabular}
\end{table*}
\normalsize

\subsection{Radio flux densities}

Table \ref{tabflux} gives the radio flux densities at 6 and 3~cm.

If a Gaussian model provided a satisfactory fit to the surface
brightness the total Gaussian flux density was adopted. This was
mostly the case for small, unresolved PNe.  If the PN was extended the
intensities within the 2 or 3 $\sigma$ level contour were summed.
This value was compared with the statistics over a larger region
across the nebula to obtain an error estimate (Fomalont, 1989).

For JaSt~69 only a small 3~$\sigma$ blob at the right position
indicated the presence of a PN. No flux or size values could be
determined.  In some cases several blobs indicated the presence of a
PN at 3~cm and hence their flux is a lower limit.  Some objects
detected at 6~cm, but having low peak flux density per beam, 
could not be seen at 3~cm.


PNe are normally optically thin at 6~cm in which case its 3~cm flux
density is about 95\% of the 6~cm flux density. In our case this means
that the flux densities are similar within the error-bars.  However
when a PN is already well resolved at 6~cm, more flux may have been
missed at 3~cm where the beam is half the size, especially if the
nebula is extended and of low surface brightness.  It is clear that
due to these factors the flux density at 3~cm is generally less
well-determined than at 6~cm, except for the bright and compact
PNe. When a PN is optically thick at 6~cm, its 3~cm flux density is
expected to be three times the flux density at 6~cm. There are some PNe like
JaSt~65 and JaSt~79 for which the 3~cm flux density is clearly higher
than the 6~cm flux density and which are point sources.  In this case
the PNe may not yet be completely optically thin at 6~cm and should be
quite young.

\begin{figure}
\caption{Histogram of the 6~cm  flux values of previously known galactic bulge PNe with
radio data and the new galactic bulge PNe (striped)}
\center{\epsfxsize=12cm \epsfbox{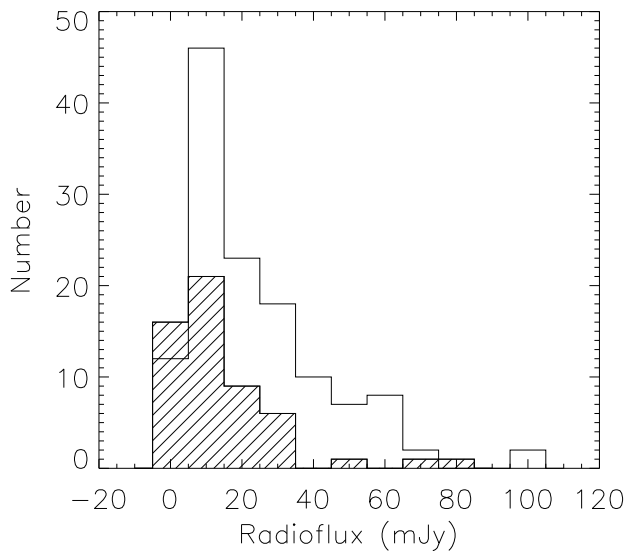}}
\label{fluxhist}
\end{figure}

In Fig. \ref{fluxhist} we plotted the histogram of the radio flux
 of the known and new galactic bulge PNe. 
We selected a sample of known galactic bulge PNe for which radio flux
densities and angular diameters are available, as in Van de Steene \&
Zijlstra (\cite{VdSteene95}).  The new bulge PNe are within 2 degrees
of the galactic center.  None has a radio flux larger than 100 mJy and
their angular sizes are smaller than 20\arcsec. Hence they fulfill the
same selection criteria as these previously known bulge PNe.

There is a larger number of PNe with low flux densities among the new
bulge PNe than among the known ones.  67~\% of the new PNe
have a radio flux less than 15~mJy, while this is only 45~\% for the
known ones.  Of the 7 known ones within 2 degrees of the galactic
center only 2 have a radio flux below 15~mJy.  The median flux for the
new PNe is 11.3~mJy, while the median for the known bulge PNe is
17.0~mJy.  Our rms noise level in the maps is similar to the
1~$\sigma$ noise of 0.1~mJy in the 6~cm maps of Zijlstra et
al. (\cite{Zijlstra89}).  Apparently these faint and small PNe have
just been missed in optical surveys done to date.

\subsection{Angular Size}

Table \ref{tabflux} also gives the angular size of the detected PNe.
We chose to determine the angular size at 6~cm because at this
wavelength the resolution was lower and thus gives the best signal to
noise ratio for the extended nebulae.  The diameter at 3~cm is given
if it is better determined than at 6~cm, such as for very small and
bright PNe.

Diameters may differ considerably depending upon how they are
calculated.  The diameter was derived by one of several ways depending
on the structure of the brightness distribution.  If a two-dimensional
Gaussian fit provided a satisfactory model to the observed
structureless surface distribution, its deconvolved FWHM major and
minor axis are given.  The equivalent diameter is the square root of
their product.  To obtain the full diameter, this value must be
multiplied with a conversion factor which is a function of the beam
FWHM and depends upon the intrinsic surface distribution of the
source.  We assumed a spherical constant emissivity shell of 0.5 and
used formula 5 and Table 1 from van Hoof (\cite{vHoof00}) to estimate
the true radii.  For small objects, if the Gaussian deconvolution was
well determined and similar at 3 and 6~cm, the FWHM at 3~cm is given.
If the deconvolution produced a point source at 6~cm, the source size
at 3~cm is given. If the source was still a point source at 3~cm, the
beam-size is an upper limit.  If the source was extended, a Gaussian
model was usually not a good representation of the radio source.  The
diameter of the PN was measured on the contour at 50~\% of the peak
and deconvolved with the beam size. To determine the full diameter we
determined the ratio of the flux density within the 3~$\sigma$ contour
with the flux density within the 50~\% contour. Hence, we assumed that
the flux decreased linearly with radius outside the 50~\% contour. We
checked that this procedure seemed to give very good agreement with
the size measured based on the 3~$\sigma$ contour.  We didn't use the
contour level at 10\% of the peak (Zijlstra et al. \cite{Zijlstra89},
Aaquist \& Kwok \cite {Aaquist90}), because this was often below noise
level.

It was noticed in the review paper by Pottasch (\cite{Pottasch92})
that there is a selection against discovering both large and small PNe
in the galactic bulge.  36~\% of the new PNe have a diameter smaller
than 5\arcsec\ , while this is 71~\% for the known ones. 
The median is 6\farcs6 for the new PNe and only 3\farcs2 for the
known bulge PNe with radio data. We seem to
identify relatively more larger PNe than in previous surveys.  In
regions with large extinction the [S\,III]$\lambda$9532 line appears
efficient in picking out the larger, low surface brightness PNe, and
not only the small and dusty ones.  Obviously these are the PNe which
may have been missed in optical surveys.

\begin{figure}
\caption{Histogram of the size of the known galactic bulge PNe with
radio data and the new galactic bulge PNe (striped)}
\center{\epsfxsize=12cm \epsfbox{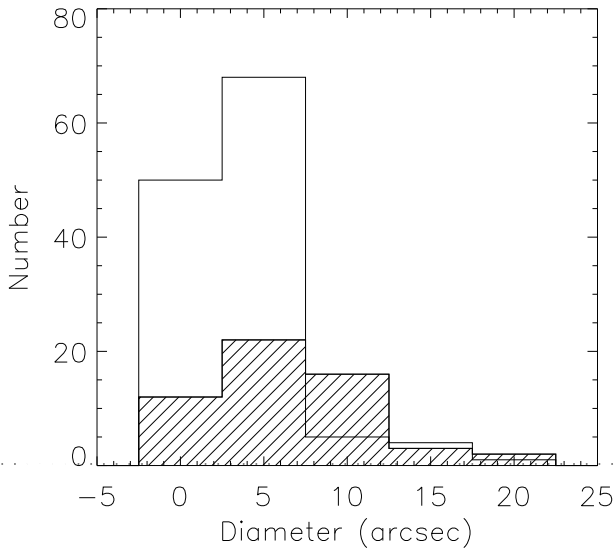}}
\label{diamhist}
\end{figure}


\section{Distances}
\label{distances} 

\begin{figure}
\caption{Histogram showing the distribution of the distances of the 
new galactic bulge PNe (striped) and the known bulge PNe with radio data.}
\center{\epsfxsize=12cm \epsfbox{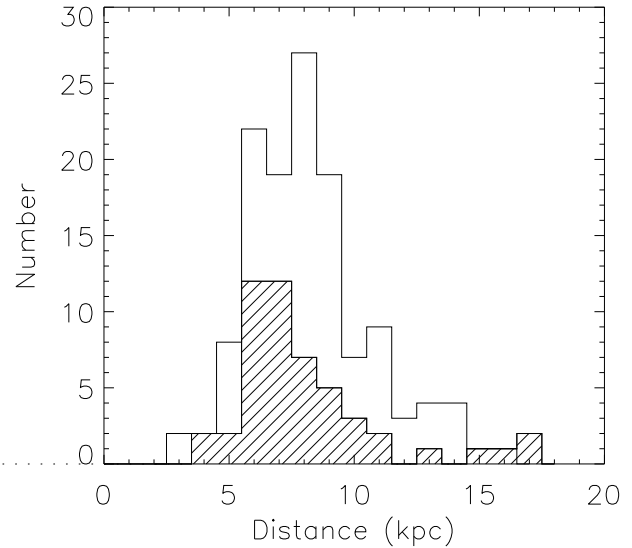}}
\label{disthist}
\end{figure}

\begin{figure}
\caption{Plot of the scale-height versus distance of the new galactic bulge
PNe (diamonds) and the known galactic bulge PNe with radio data (crosses)}
\center{\epsfxsize=12cm \epsfbox{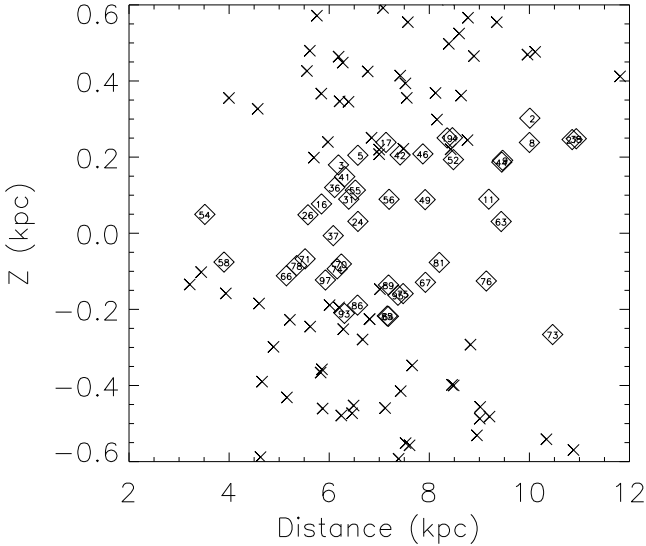}}
\label{distsch}
\end{figure}

Based on the flux densities and the angular equivalent diameters we calculated
the statistical distances to the PNe as established in Van de Steene
\& Zijlstra (\cite{VdSteene95}).  The relation is not strictly valid for
optically thick and very small nebulae 
 which  may have their distance overestimated.
The distances will tend to be overestimated for PNe with lower surface
brightness than average, while PNe with a higher surface brightness
than average will tend to have their distances underestimated.
However, statistically the distance distribution will be representative.



The distances are presented in Table \ref{quantities}.
A histogram of the distance distribution is
presented in Fig. \ref{disthist}. The median distance is 7.2 kpc,  
about 8\% closer than the galactic center; presumably, the more distant
PNe suffer greater extinction and fell below our detection limit in our survey.

In Fig. \ref{distsch} we plotted the scale height versus the distance.
It seems that we observe PNe at the edge of the galactic bulge. As we
look further inward, the extinction probably becomes too large.  The
PNe further away are at larger scale height where, apparently, the
extinction is less. The known PNe are generally further away from the
galactic center, surrounding our new PNe. The median absolute scale
height is 136 pc.

\begin{table}
\caption[]{In this table we present the statistical distance  in
column 2, the scale height, radius, and ionized mass calculated
using this distance in columns 3, 4, and 5 respectively, and
E(B-V)~$=$~c$_{H_{\beta}}$~/~1.46 in column 6. If the object was 
not detected in H$\alpha,$ E(B-V) is marked with X. If no H$\alpha$
image was obtained E(B-V) is marked with Z. In this case the E(B-V) value
was derived from its spectrum. Uncertain values,
due to uncertain radio flux or H$\alpha$ flux values,
are marked with a colon. PS stands for Point Source. }
\begin{tabular}{lllllll}
JaSt & Dist & z  & R  & M$_{ion}$ & E(B-V) &  Comment \\
     &  kpc & pc & pc &  M$_{\sol}$ &   &       \\
\hline
1  & 14.6: & 401.6:  & 0.20: & 0.25: & 2.5: & \\
2  & 10.0 & 305.7  & 0.15 & 0.19 & 2.0 & \\
3  & 6.2  & 183.0   & 0.14 &  0.17 & 2.3 & \\
4  & 8.5  & 258.9   & 0.23 &  0.29 & 1.9 & \\
5  & 6.6  & 210.4   & 0.13 &  0.17 & 2.3 & \\
8  & 10.0 & 241.9  & 0.17 &  0.22 & 2.1 & \\
9  & 9.5   & 194.9  & 0.19 &  0.23 & 2.9 & \\
11 & 9.2  &  91.6   & 0.24 &  0.31 & 3.7 & \\
16 & 5.9  &  77.6   & 0.07 &  0.08 & 3.5 & \\
17 & 7.1  & 242.5  & 0.14 &  0.17 & 1.8 & \\ 
19 & 8.4  & 264.7  & 0.14 & 0.17 & 1.8 & \\
23 &  &             &  &    &  2.4 & PS \\
24 & 6.6  &  31.4  & 0.09 & 0.11 & X & \\
26 & 5.6 &   49.7  & 0.15 & 0.19 & X & \\
27 & 10.8 & 246.2 & 0.23 & 0.29 & 2.3 & \\
31 & 6.4  &  91.2  & 0.14 & 0.17 & 3.7 & \\
34 & 13.2 & 392.9 & 0.23 & 0.29  & 2.1 & \\
36 & 6.1  & 120.2  & 0.05 & 0.06 & 2.7 & \\
37 & 6.1  &  -6.3  & 0.11 & 0.13  & X & \\
38 & 10.9: & 251.7:  & 0.19: & 0.23: & 2.0: & \\
41 & 6.3  & 150.4  & 0.10 & 0.12  & 2.1 & \\
42 & 7.4  & 207.0  & 0.09 & 0.11 & 2.1 & \\
44 & 9.6:  &  190.2: & 0.20: & 0.26: & 1.9: & \\
46 & 7.9  &  207.1 & 0.05 & 0.06 & 1.9 & \\
49 & 7.9  &  89.4 & 0.13 & 0.16  & X  &  \\ 
52 & 8.5  &  190.6 & 0.03 & 0.04 & 2.1 & \\
54 & 3.5  &   50.2   & 0.06 & 0.07 & X & \\
55 & 6.5  &  115.9 & 0.17 & 0.21 & 2.5 & \\
56 & 7.2   &  89.9   & 0.09 & 0.11& X &  \\ 
58 & 3.9   &  -77.8  & 0.15 & 0.18 & 4.1&  \\
60 & 16.2  &  -213.6  & 0.04 & 0.04 & X & \\
63 & 9.4  &  31.5  & 0.19 & 0.25 & 2.6 & \\
64 & 7.2  & -216.3 & 0.04 & 0.05 & 2.6 & \\
65 &        &            &          &          & 1.3 & PS \\
66 & 5.5  &  -111.1 & 0.03  & 0.04 & 2.8 & \\
67 & 7.9  &  -127.0 &  0.03 & 0.04 & 4.6: Z & \\
68 & 16.7 &  -388.0 &  0.04 & 0.04 & 2.8 & \\
70 & 6.2   &  -82.0   & 0.13 & 0.16 & X & \\
71 & 5.5 &  -67.4 & 0.06 & 0.07 & 3.6 & \\
73 & 10.5  & -261.7 &  0.04 & 0.05 & 1.9 &  \\
74 & 6.2  &  -93.9 & 0.07 & 0.09  & 3.6 & \\
75 & 7.5  & -157.5 & 0.05 & 0.06 & 2.7 & \\
76 & 16.8 &   -223.1 & 0.03 & 0.03 & 5.0: Z & \\
77 &  &  &  &    & 2.4   & PS \\
78 & 5.4  &  -87.5 & 0.19 & 0.24 & 3.4: Z & \\
79 & & &  &  & 1.5 & PS \\
81 & 8.2  &  -75.8 & 0.04 & 0.05 &  X & \\
85 & 7.2:   &  -224.1 & 0.30 & 0.39: & 1.2 & \\
\hline
\end{tabular}
\label{quantities}
\end{table}
\addtocounter{table}{-1}
\begin{table}
\caption[]
{ Table \ref{quantities} continued}
\begin{tabular}{lllllll}
JaSt & Dist & z  & R  & M$_{ion}$ & E(B-V) &  Comment \\
     &  kpc & pc & pc &  M$_{\sol}$ &   &       \\
\hline
86 & 6.6  & -193.0 & 0.18 & 0.23 & 1.9 & \\
89 & 7.2  & -138.3 & 0.15 & 0.19 & 2.1 & \\
90 & 17.2: &  -303.3: & 0.23: & 0.29: & 2.1 & \\
93 & 6.5:  & -217.2: & 0.27: & 0.35: & 1.9: Z  & \\
95 & 7.4:  & -157.7: & 0.23: & 0.28: & 1.3 & \\
97 & 5.9  & -125.5 & 0.16 & 0.21 & 2.6 & \\
98 &        &             &         &       &  2.7 Z & PS \\
\hline
\end{tabular}
\end{table}

\section{Extinction}
\label{extinction}

\begin{figure}
\caption{Longitude versus latitude of the new galactic bulge PNe detected in
the radio.
The numbers inside the symbols indicate their E(B-V) value.}
\center{\epsfxsize=12cm \epsfbox{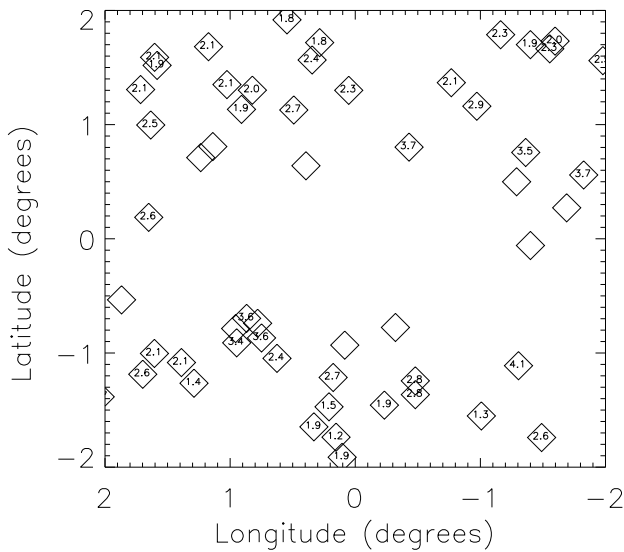}}
\label{ext_lb}
\end{figure}

\begin{figure}
\caption{Distance versus scale height of the new galactic bulge PNe detected in
the radio. The numbers inside the symbols indicate their E(B-V) value.}
\center{\epsfxsize=12cm \epsfbox{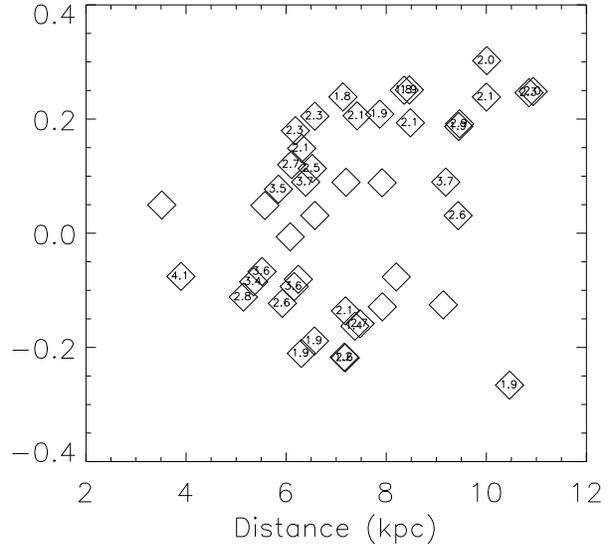}}
\label{ext_schdist}
\end{figure}

We calculated the extinction E(B-V) comparing the radio flux densities
with the total H$\alpha$ flux values, which were obtained from images
obtained with the NOAO 8k~x~8k Mosaic Imager on the 0.9-m telescope at KPNO
(Jacoby
\& Van de Steene  2001, in preparation). The flux values were corrected for the
contribution of [N II]$\lambda$6548~\&~$\lambda$6584
 based on spectra obtained at ESO and CTIO (Van de Steene \& Jacoby,
2001, in preparation).  The
H$\beta$ flux and radio flux density have the same dependency on electron density
and the expected ratio is only a weak function of the electron
temperature and helium abundance (Pottasch, \cite{Pottasch84}).  
The radio flux density and H$\beta$ flux can be used to determine the
extinction c$_{H\beta}$~$=$~1.46~E(B-V). Assuming the standard ratio of
H$\alpha$ / H$\beta$~$=$~2.85, we used the H$\alpha$ flux to
predict the extinction: c$_{H\alpha}$ = E(B-V). The values of E(B-V)
are presented in Table \ref{quantities}.  All but one PN have values between
1 and 4. For the PNe detected in the radio but not in H$\alpha$ the
extinction is likely to be higher.

As mentioned before: some PNe were detected in H$\alpha$ but not in
the radio, probably due to too low surface brightness. 
Eight PNe were detected in the radio, but not in the H$\alpha$ images,
probably due to too high extinction. These PNe also show no
H$\alpha$ emission in their spectra. They are visible at longer wavelengths such
as [S~III]$\lambda$9532.

From Fig. \ref{ext_lb} and \ref{ext_schdist} it appears indeed that the extinction
 increases towards the galactic center. No extinction value could
be determined for objects closest to the galactic center.
The distances of JaSt 54 and JaSt 58  appear to be underestimated,
unless their internal extinction is very large.

\section{Conclusions}

We obtained the radio flux densities and diameters for 64 new galactic bulge
PNe with the Australia Telescope Compact Array.
\begin{enumerate}
\item  We have
a larger ratio of larger PNe with angular diameters
around 10\arcsec\ and radio flux densities below 15~mJy, than in previous surveys.  
Our survey seems to have picked out the larger low surface brightness PNe
which were missed in optical surveys to date.
\item We calculated their distances according to the method presented in Van de Steene \&
Zijlstra (\cite{VdSteene95}).  The new galactic bulge PNe are mainly located around
the galactic center, and closer to it than the previously known PNe. 
\item We calculated the E(B-V) extinction value based on the radio flux densities
and the total H$\alpha$ flux values from imaging.
Only 1 PN was detected with a E(B-V) value larger than 4.
Generally speaking, when \\ E(B-V)~$>$~4.0 the galactic bulge PNe become undetectable
in H$\alpha$.
\end{enumerate}

\appendix
\section{Contour Plots: available upon request from gsteene@oma.be}


\begin{thebibliography}{}

\bibitem[1990]{Aaquist90}
Aaquist, A., Kwok, S., 1990, A\&AS 84, 229 

\bibitem[1992]{Acker92}
Acker, A., Marcout, J., Ochsenbein, F., Stenholm, Tylenda, R., 1992,
Strasbourg-ESO catalogue of galactic planetary nebulae, ESO 

\bibitem[1997]{Dopita97}
Dopita, M.A., Vassiliadis, E., Wood, P.R., Meatheringham, S.J., Harrington, J.P., Bohlin, R.C., Ford, H.C., Stecher, T.P., Maran, S.P., 1997, ApJ 474, 188 

\bibitem[1989]{Fomalont89}
Fomalont, E., Synthesis imaging in radio astronomy, 1989, PASP 6, ed. A. Perley,
F.R. Schwab, and A. Bridle, p. 222 

\bibitem[1994]{Kohoutek94}
Kohoutek, L., 1994, Astron. Nachr. 315, 235 

\bibitem[1984]{Pottasch84}
Pottasch, S.R., 1984, "Planetary Nebulae", D. Reidel (Dordrecht) , p.93, p.97

\bibitem[1989]{Pottasch89}
Pottasch, S.R., Acker a., 1989, A\&A 221,123 

\bibitem[1992]{Pottasch92}
Pottasch,S.R., 1992, A\&ARv 4, 256 

\bibitem[1995]{Sault95}
Sault R.J., Teuben P.J., Wright M.C.H., 1995, ``A retrospective of Miriad'',
Astronomical Data Analysis Softwared and Systems IV, ed. R. Shaw, H.E. Payne, 
J.J.E. Hayes, ASP Conf. Ser., 77, 433 

\bibitem[1995]{VdSteene95}
Van de Steene, G.C., Zijlstra, A.A., 1995, A\&A, 293, 541

\bibitem[1999]{vHoof99}
van Hoof, P.A.M., Van de Steene, G.C., 1999, MNRAS, 308, 623  

\bibitem[2000]{vHoof00}
van Hoof, P.A.M., 2000, MNRAS, 314, 99

\bibitem[2000]{Walsh00}
Walsh, J.R., Jacoby, G.H., Peletier, R.F., Walton, N.A., 2000,
SPIE, Discoveries and Research Prospects from 8 to 10-Meter
Class Telescopes, Vol 4005, p. 131 

\bibitem[1989]{Zijlstra89}
Zijlstra, A.A., Pottasch, S.R., Bignell, C., 1989, A\&AS, 79, 329

\end{thebibliography}
\end{document}